# Improving Lesion Segmentation in FDG-18 Whole-Body PET/CT scans using Multilabel approach: AutoPET II challenge

Gowtham Krishnan Murugesan, Diana McCrumb, Eric Brunner, Jithendra Kumar, Rahul Soni, Vasily Grigorash, Stephen Moore, and Jeff Van Oss

**Abstract:**

Automatic segmentation of lesions in FDG-18 Whole Body (WB) PET/CT scans using deep learning models is instrumental for determining treatment response, optimizing dosimetry, and advancing theranostic applications in oncology. However, the presence of organs with elevated radiotracer uptake, such as the liver, spleen, brain, and bladder, often leads to challenges, as these regions are often misidentified as lesions by deep learning models. To address this issue, we propose a novel approach of segmenting both organs and lesions, aiming to enhance the performance of automatic lesion segmentation methods. In this study, we assessed the effectiveness of our proposed method using the AutoPET II challenge dataset, which comprises 1014 subjects. We evaluated the impact of inclusion of additional labels and data in the segmentation performance of the model. In addition to the expert-annotated lesion labels, we introduced eight additional labels for organs, including the liver, kidneys, urinary bladder, spleen, lung, brain, heart, and stomach. These labels were integrated into the dataset, and a 3D UNET model was trained within the nnUNet framework. Our results demonstrate that our method achieved the top ranking in the held-out test dataset, underscoring the potential of this approach to significantly improve lesion segmentation accuracy in FDG-18 Whole-Body PET/CT scans, ultimately benefiting cancer patients and advancing clinical practice.

**Introduction:**

Whole-body Positron Emission Tomography/Computed Tomography (PET/CT) is a vital tool for imaging tumors, aiding in early detection of metastatic lesions, quantifying metabolically active tumors, and contributing significantly to cancer diagnosis, staging, treatment planning, and recurrence monitoring. Recent studies suggest its potential for personalized therapy response assessment. Among PET radiotracers, 18-Fluorodeoxyglucose (18F-FDG) is the mostly used due to its ability to target increased glucose metabolism in malignant tumors, demonstrating high sensitivity for metastasis detection in solid tumors. [1–5].

Simultaneously, the evolution of deep learning algorithms has brought about a revolution in the field of medical imaging, fostering more precise and efficient segmentation of cancerous lesions within acquired images[6–9]. These algorithms possess the remarkable ability to autonomously delineate tumor boundaries, even in scenarios where conventional methodologies fail due to the intricacy, variability, or subtlety of the lesions. The integration of deep learning algorithms into the molecular theranostics framework holds immense promise, streamlining the interpretation of diagnostic data and optimizing subsequent therapeutic strategies. This includes precise radiotracer dose estimation, treatment planning by monitoring treatment response, and the staging of cancer.

Radiotracer uptake exhibits interpatient variability, and the inherent design of radiotracers may lead to heightened accumulation in normal organs with high metabolic activity, such as the brain, or cleansing organs like the liver, kidneys, and urinary bladder. Automatic lesion segmentation algorithms often struggle to differentiate between normal uptakes in these organs and actual lesions, presenting a significant challenge. Consequently, we hypothesize that segmenting high-uptake organs alongside lesions can provide the model with valuable discernment between the two. Our preliminary study[10], training a multiclass model to segment lesions in 40 prostate cancer subjects with WB Ga-68 PSMA-11 PET/CT scans and testing on 10 held-out subjects, revealed that the proposed multilabel model significantly outperformed the single-label model, which focused solely on lesion segmentation.

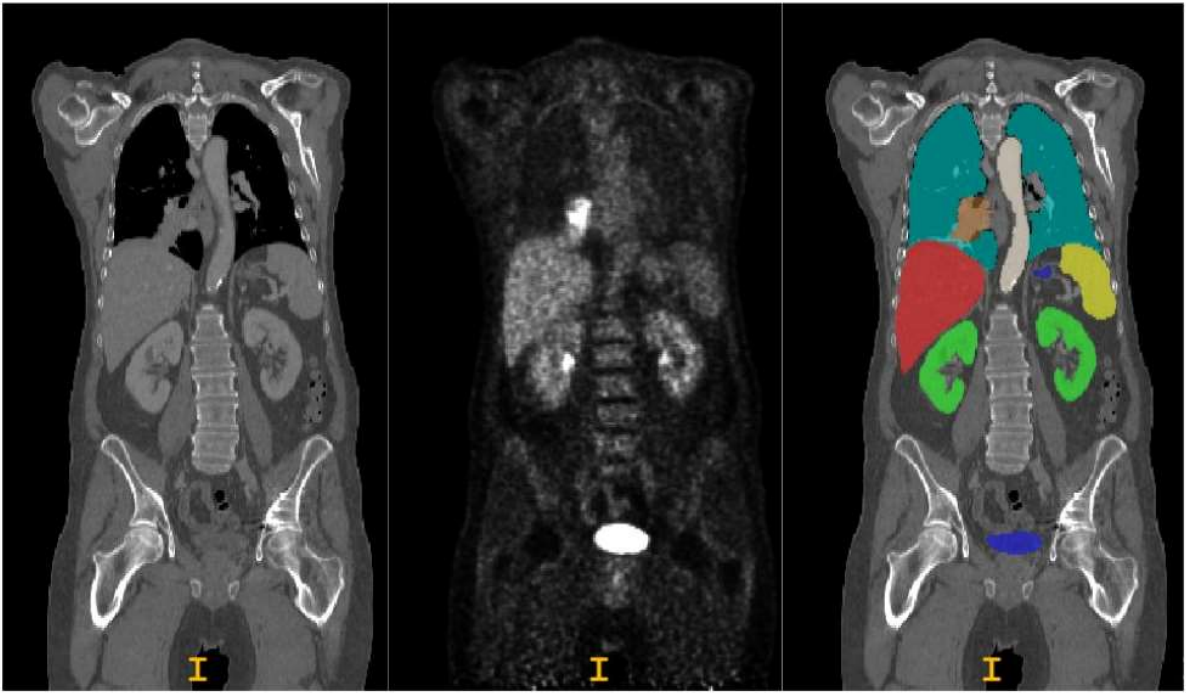

*Figure 1:* **Training data**: *Representation of the Whole-Body CT (left), PET (middle), and labels (right) employed to train the final model.*

In this research paper, we extend our original investigation by including a larger cohort from the AutoPET challenge. We leveraged a 3D UNet model within the nnUNET framework and employed a rigorous 5-fold cross-validation approach for evaluation. To assess our method, we initially conducted experiments by splitting the training data into two groups: one containing 819 studies and the other 195 studies, ensuring that studies from the same subjects were kept together in each group. We conducted studies to evaluate the effect of adding more labels and more data in the model's performance in segmenting lesions.

Subsequently, we trained a final model using a 5-fold cross-validation technique to segment lesions and other high-uptake organs, incorporating all 1014 studies from the AutoPET training dataset. An ensemble of the final model across all five folds was created and submitted to the AutoPET challenge for benchmarking against other methods. This comprehensive methodology and evaluation process allowed us to assess the performance of our approach within the context of the AutoPET challenge and make comparisons with other submitted techniques. Significantly, our proposed method achieved the top-ranking position in the AutoPET II challenge, highlighting its potential as a clinical utility in the field of molecular theranostics.

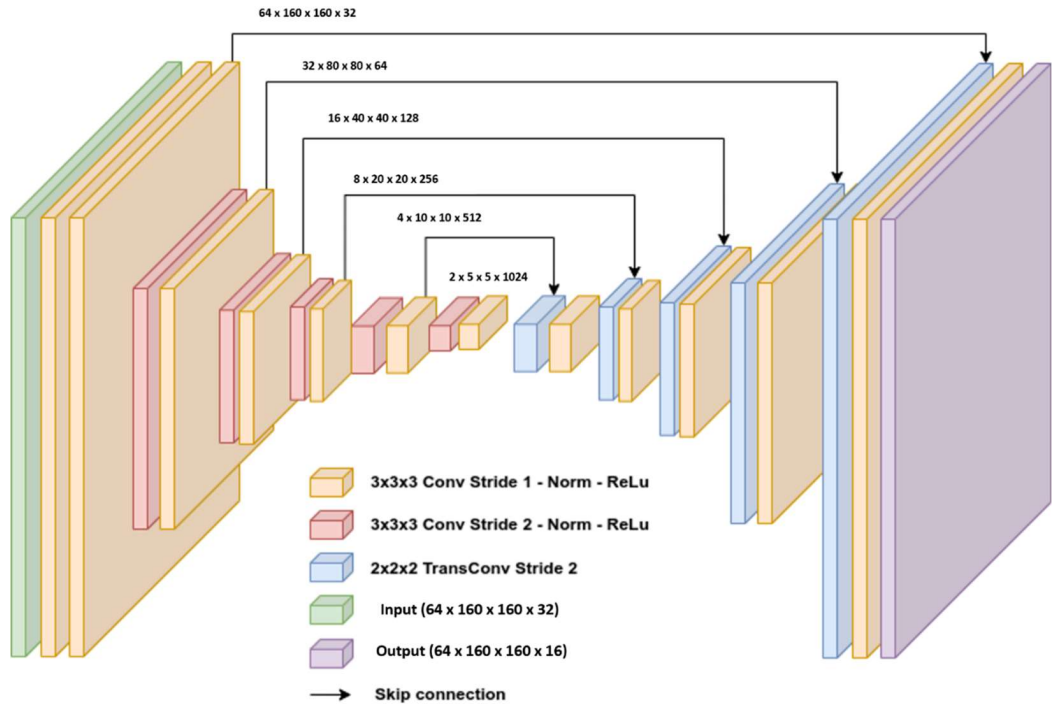

*Figure 2: Figure 2: The layers of the UNET architecture used. The input is a volume of 128x128x128 with one channel, CT. Input is resampled down five times by convolution blocks with strides of 2. On the decoder side, skip connections are used to concatenate the corresponding encoder layers to preserve spatial information.*

**Methods:**

**Data and Preprocessing**

The models were trained using whole-body FDG-PET/CT data from 900 patients, including 1014 studies provided by the AutoPET challenge II 2023[11,12]. A held-out dataset consisting of 150 studies, 100 of which originated from the same hospital as the training database and 50 were drawn from a different hospital with a similar acquisition protocol, was used as a test dataset to assess the robustness and generalizability of the algorithm. In the preprocessing step, the CT data was resampled to the PET resolution and normalized. Two experts annotated the training and test data. A radiologist with 10 years of experience in Hybrid Imaging and experience in machine learning research at the University Hospital Tübingen annotated all data. A radiologist with 5 years of experience in Hybrid Imaging and experience in machine learning research at the University Hospital of the LMU in Munich annotated all data. We randomly split the training data into 5-fold and trained the 3D UNet model within nnUNet framework to segment multiple organs and lesions. The final model is the ensemble of the five folds and is uploaded into the challenge portal for testing in docker format.

**Effect of adding multiple labels and more data:**

To assess our method, we initially conducted experiments by splitting the training data into two groups: one containing 819 studies and the other 195 studies, ensuring that studies from the same subjects were kept together in each group. We conducted studies to evaluate the effect of adding more labels (liver, spleen, kidneys, urinary bladder) and more data in the model's performance in segmenting lesions.

We conducted a thorough model training and evaluation process utilizing a 5-fold cross-validation strategy, with a specific focus on two distinct 3D UNET models designed for lesion segmentation. One model was tailored for isolating lesions (single label), while the other was created for segmenting lesions in conjunction with other high uptake organs (multi label). The labels of high uptake organs are derived using openly available totalsegmentator[13]. We initially trained both models on the dataset comprising 819 studies and rigorously assessed their performance on a separate dataset of 195 distinct medical studies. This evaluation allowed us to gauge the models' effectiveness in a real-world context.

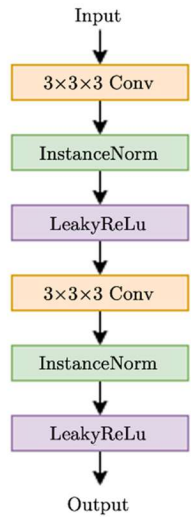

*Figure 3: In our UNET models, each encoding layer is a series of Convolution, normalization, and activation function repeated twice.*

To investigate the impact of dataset size on model performance, we conducted a second phase of experimentation. We randomly selected a subset of 100 studies from the original 819 and trained two separate models with identical segmentation objectives—lesion-only segmentation and joint lesion and high uptake organ segmentation. These models were then evaluated using the same 195-study held-out dataset, enabling us to understand how dataset size influenced their segmentation capabilities. This systematic approach provided valuable insights into the models' suitability for various practical applications in medical image analysis.

**AutoPET II Challenge**

For the challenge submission we trained a multilabel model using all the 1014 studies from AutoPET II challenge training data. We derived eight additional organs (eight organs: the liver, kidneys, urinary bladder, spleen, lung, brain, heart, and stomach) using openly available totalsegmentator[13] package and added to the training dataset (Fig.1). The final model is the ensemble of the five folds and is uploaded into the challenge portal for testing in docker format.

**Model Training Methodology**

**Model Architecture**:

The nnUNET pipeline has consistently demonstrated top-tier performance in various medical imaging segmentation competitions.[14]. The standard variant 3D UNet model was utilized, incorporating skip connections for training, as illustrated in Figures 3 and 4. The input image dimensions are set to 128x128x128 with a single channel, utilizing CT scans as input data. The input is progressively downsampled by convolution blocks with 2x strides through five stages. On the decoder side, skip connections are utilized to combine corresponding encoder layers, preserving spatial information. Network layers incorporate instance normalization and utilize leaky ReLU activation. The initial architecture employs 32 feature maps, which double during each downsample operation in the encoder, reaching up to 1024 feature maps before halving again during transposed convolutions in the decoder. The decoder output maintains the same spatial dimensions as the input, followed by a 1x1x1 convolution to produce a single-channel output, which is processed with a SoftMax function. During training, models are trained over five folds using a loss function combining the Dice Sorensen Coefficient (DSC) and weighted cross-entropy loss to mitigate overfitting. Augmentation techniques such as random rotations, random scaling, random elastic deformations, gamma correction, mirroring, and elastic deformations are employed to enhance model robustness. Each of the five models is trained for 1000 epochs with a batch size of eight, utilizing the SGD optimizer with a learning rate of 0.01. Performance assessment includes metrics such as Dice Similarity Coefficient (DSC) and normalized surface dice (NSD) to evaluate various aspects of the segmentation methods.

**Results:**

**Effect of adding multiple labels and more data:**

The model trained on a subset of 100 subjects achieved Dice Similarity Coefficients (DSC) of 0.63+/-0.04 (95% confidence interval (CI): 0.57,0.69), FPV of 1330.45 (95% CI: 165.08,256.02), FNV of 210.55 (95% CI: 98.35,141.88) for single-label segmentation and 0.78+/-0.02 (95% CI: 0.7,0.80) , FPV of 351.13 (161.40, 233.49) and FNV of 197.44 (95% CI: 98.35, 141.88)for multi-label segmentation. In contrast, models trained using the entire dataset of 819 studies achieved higher DSC scores, specifically 0.66+/-0.08 (95% CI: 0.54, 0.77),FPV of 1456.72 (95% CI: 262.03, 279.87), and FNV of 165.30 (95% CI: 98.37, 141.88) for single-label segmentation and 0.79+/-0.02 (95% CI: 0.77,0.82) , FPV of 389.28 (95% CI: 295.83, 482.72) and FNV of 120.12 (95% CI: 98.35, 141.88) for multi-label segmentation (Fig.4 and 5 )(Table.1).

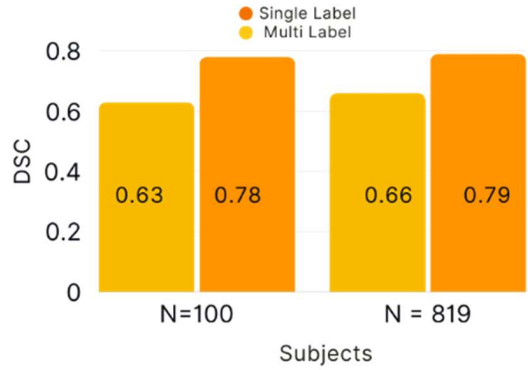

*Figure 4: **Comparison of Dice Coefficients** between single label (yellow) and multilabel (orange) models in small (N=100) and large (N=819) cohorts*

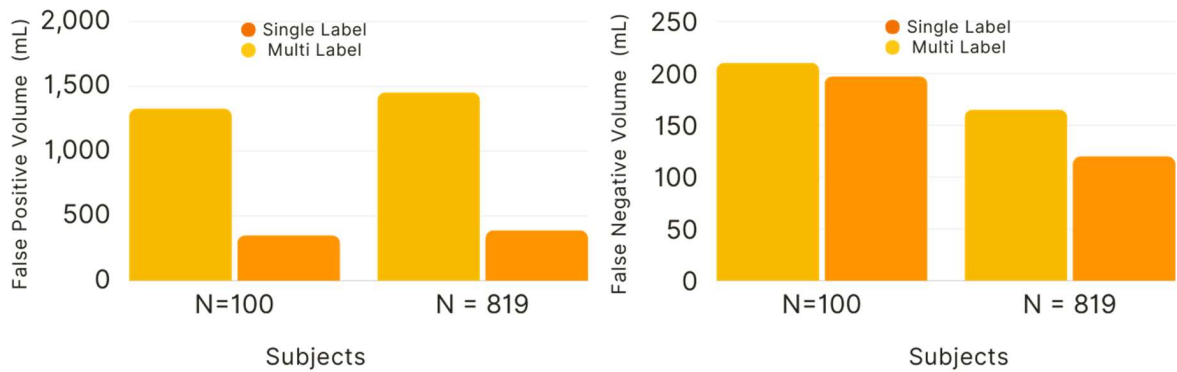

*Figure 5: Comparison of false positive volumes (left) and false negative volumes (right) between single label (yellow) and multilabel (orange) models in small (N=100) and large (N=819) cohorts*

*Table 1: Dice, FPV and FNV evaluation for single label and multilabel models with larger and smaller cohorts. 95% confidence interval metrics are show within parentheses.*

|  | Singlelabel | | Multilabel | |
|---|---|---|---|---|
|  | N=100 | N=819 | N=100 | N=819 |
| Dice | 0.63 (0.57,0.69) | 0.66 (0.54,0.77) | 0.78 (0.70, 0.80) | 0.79 (0.77, 0.82) |
| FPV | 1330.45 (165.08,256.02) | 1453.72 (262.03, 279.87) | 351.13 (161.40, 233.49) | 389.28 (295.83, 482.72) |
| FNV | 210.55 (98.35, 141.88) | 165.3036 (98.37, 141.88) | 197.44 (98.35, 141.88) | 120.12 (9.35, 141.88) |

**AutoPET II Challenge**

Our method achieved top ranking in AutoPET II challenge by achieving xx dice, yy FPV and zz FNV.

**Discussion and Conclusion:**

In this study, we conducted training and evaluation of two distinct deep learning models aimed at segmenting lesions within medical imaging data. One model was designed to solely focus on lesion segmentation, while the other was equipped to segment lesions as well as other anatomical structures. We conducted this analysis using two separate datasets, one of smaller cohort comprising 100 cases (N=100) and another of larger cohort encompassing 819 cases (N=819). Our results revealed a clear advantage for the multi-label model when compared to the single-label model. The multi-label model demonstrated superior performance in terms of Dice scores, while also exhibiting lower False Positive Volumes (FPV) and False Negative Volumes (FNV).

The outcomes of this study provide compelling support for our initial hypothesis, which posited that incorporating high radiotracer uptake regions in addition to lesions would yield improved lesion segmentation performance. It is noteworthy that the introduction of multiple labels for training significantly enhanced the performance of the model trained on the smaller dataset, achieving performance levels comparable to the multi-label model trained on the larger dataset. Further, the multi-label model trained on the smaller dataset outperformed the single-label model trained on the larger dataset. This underscores the significance of incorporating multiple labels, not just additional data, in enhancing the lesion segmentation capabilities of deep learning models. Moreover, our proposed method secured the top ranking in the AutoPET II challenge, serving as strong evidence for the broad applicability and effectiveness of our approach.

Furthermore, our findings emphasize the versatility and effectiveness of the proposed approach. They underscore its capacity to enhance lesion segmentation performance, even in scenarios where the dataset is limited in size, an attribute of paramount importance in the realm of medical imaging data analysis. In conclusion, our research demonstrates that the inclusion of high radiotracer uptake regions and other anatomical structures as multiple labels significantly enhances lesion segmentation performance, even when working with smaller datasets.